\documentclass[12pt]{article}
\usepackage[final]{epsfig}
\usepackage{graphics}
\usepackage{amsmath}
\usepackage{amsfonts}
\usepackage{cancel}
\usepackage{latexsym}
\usepackage{amssymb}
\usepackage{graphicx}
\usepackage{epstopdf}
\usepackage{url}
\usepackage{esvect}
\usepackage{graphicx,epsfig,psfrag}

\newtheorem{thm}{Theorem}

\newtheorem{rem}{Remark}

\begin{document}
\newcommand{\eps}{{\varepsilon}}
\newcommand{\proofend}{$\Box$\bigskip}
\newcommand{\C}{{\mathbf C}}
\newcommand{\Q}{{\mathbf Q}}
\newcommand{\R}{{\mathbf R}}
\newcommand{\Z}{{\mathbf Z}}
\newcommand{\RP}{{\mathbf {RP}}}
\newcommand{\CP}{{\mathbf {CP}}}
\newcommand{\A}{{\rm Area}}
\newcommand{\Le}{{\rm Length}}


\newcommand{\marginnote}[1]
{
}
\newcounter{ml}
\newcommand{\bk}[1]
{\stepcounter{ml}$^{\bf ML\thebk}$%
\footnotetext{\hspace{-3.7mm}$^{\blacksquare\!\blacksquare}$
{\bf ML\thebk:~}#1}}

\newcounter{st}
\newcommand{\st}[1]
{\stepcounter{st}$^{\bf ST\thest}$%
\footnotetext{\hspace{-3.7mm}$^{\blacksquare\!\blacksquare}$
{\bf ST\thest:~}#1}}


\title {Rotating saddle trap: A Coriolis force in an inertial frame}
\author{Oleg Kirillov\thanks{Steklov Mathematical Institute, Russian Academy of Sciences, Gubkina 8, Moscow 119991, Russia (visiting Universita di Trento, DICAM, via Mesiano, 77 I-38123 Trento,
Italy); e-mail: \tt{kirillov@mi.ras.ru}
}
\, and Mark Levi\thanks{
Department of Mathematics,
Pennsylvania State University, University Park, PA 16802, USA;
e-mail: \tt{levi@math.psu.edu}
}
\\
}

\date{\today}
\maketitle
\begin{abstract}
Particles in rotating saddle potentials exhibit precessional motion which,  up to now, has been explained by explicit computation. We show that this precession is due to a hidden Coriolis--like force which, unlike the standard Coriolis force, is present in the inertial frame. We do so by finding a hodograph--like ``guiding center" transformation using the method of normal form.
 \end{abstract}

\section{Introduction and background}
\begin{figure}[thb]
	\center{  \includegraphics{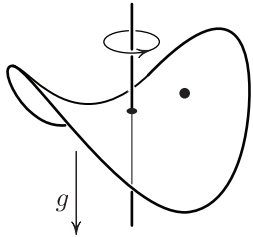}}
	\caption{A particle on a rotating saddle surface.}
	\label{fig:saddlesurface}
\end{figure}

We consider the motion of a particle in the rotating saddle potential  in the plane. The ``spinning" potential whose graph is obtained by rotating the graph of a fixed potential $ U_0(x  )=U_0(x_1,   x_2) $ with angular velocity $\omega$ is
\[
	U (x  , t) = U_0(R ^{-1}x  ), \ \   x  =(x_1,x_2) \in {\mathbb R}  ^2
\]
where
\[
	 R=R(\omega t) =
	\left( \begin{array}{cc} \cos \omega t & -\sin \omega t  \\ \sin \omega t  & \ \  \cos \omega t \end{array} \right)
\]
is the counterclockwise rotation. If $ U_0 $  is a saddle with equal principal curvatures, then
\[
	U_0(x  )=\frac{1}{2} (x_1 ^2 - x_2 ^2 )
\]
without the loss of generality, and the equations of motion $ \dot x  = - \nabla U $ take the form
\begin{equation}
	\ddot x   +   S( \omega t ) x   =0, \   x   \in {\mathbb R}  ^2,
	\label{eq:linear.trap}
\end{equation}
where
\[
	  S(\tau)= \left( \begin{array}{rr} \cos 2\tau  & \sin 2\tau  \\ \sin  2\tau & -\cos 2\tau  \end{array} \right),  \  \ \tau =
	  \omega t.
\]
These equations describe the linearized motion of a
  particle sliding without friction on  a rotating saddle surface (with equal and opposite principal curvatures) in the presence of gravity,
Figure~\ref{fig:saddlesurface}.   It is a surprising fact, known for almost a century \cite{B1918,B1975,B1976,V95,K2013a,K2013} that the equilibrium position of the particle becomes stable if  the surface   rotates around the vertical axis sufficiently fast (a heuristic explanation of this effect can be found in \cite{HB2005}, and is also given below). Numerical    experiments show another puzzling effect: the Foucault--like prograde precession happening in the inertial frame \cite{HB2005,P1970,THB2002}, Figure~\ref{fig:illustration1}.
\vskip 0.1 in
A similarly surprising phenomenon  is the stabilization of a ball rolling without slipping on a rotating saddle surface (several demonstrations can be found on YouTube) \cite{THB2002,RST1995}. Superficially, the two effects  appear to be the same; however,  the reasons for stability are   entirely different. For the rolling ball,   the gyroscopic effect, which has no counterpart for a point mass in our case, plays the key role. In fact, the rolling ball is stable even if the surface is horizontal and flat, \cite{weckesser}. The rolling ball  is an entirely different system: first, it is  a nonholonomic system (see \cite{naimark-fufaev} for more details), unlike the one considered here,  and second,  it has more degrees of freedom.
 \vskip 0.1 in
Returning to the particle in the rotating saddle potential, the  force field  $ - R(\omega t ) x $ in  (\ref{eq:linear.trap})  admits the following nice interpretation.  Consider  the
saddle force field $ \langle x, -y \rangle $,  and make it time--dependent by    rotating each vector counterclockwise  with angular velocity $ 2\omega $.
Equation (\ref{eq:linear.trap})  describes the motion of the unit point mass  in this  force field.

We note that $S(\tau)$ is   the reflection in the $x$--axis followed by a rotation by $ 2 \tau  $ counterclockwise. Equivalently,  $S$ is the reflection with respect to the line which forms the angle $ 2\tau $ with respect to the $x$--axis.
Thus $S$ is an anti--orthogonal matrix, with eigenvalues $ \pm 1 $, and satisfying
\begin{equation}
	S ^2 = I,
	\label{eq:idempotent}
\end{equation}
the identity matrix.

\section{The main result: a ``hodograph" transformation.}

The following  result was stated in \cite{KL2015}, but without the derivation.
\begin{thm}\label{mainthm}
Given a vector function $ x  : {\mathbb R}  \rightarrow {\mathbb R}  ^2  $, consider  its ``guiding center", or the ``hodograph" image
\begin{equation}
	 u=x  - \frac{\varepsilon ^2}{4} S(t/ \varepsilon)(x  - \varepsilon J \dot x   ), \  \
	 J = \left( \begin{array}{cc} 0 & -1 \\ 1& 0 \end{array} \right) .
	\label{eq:guidingcenter}
\end{equation}

If $ x  (t) $ is a solution of  (\ref{eq:linear.trap}), then $ u(t) $ satisfies
\begin{equation}
	\ddot u - \frac{\varepsilon ^3}{4}    J\dot u+ \frac{\varepsilon ^2}{4} u=
	 \varepsilon ^4 f (u, \dot  u , \varepsilon ),
	\label{eq:mainaveq1}
\end{equation}
where $ f $ is a  function linear in $  u$, $ \dot  u $ and analytic in $\varepsilon$,   in a fixed neighborhood of   $ \varepsilon =0  $. The guiding center therefore behaves,
 (ignoring the $O(\varepsilon ^4) $--terms)  as a point   charge of unit mass in the  potential $ V( u) = \frac{\varepsilon ^2}{8}  u ^2 $  in the magnetic   field of constant magnitude
 $ B=\varepsilon ^3/4$   perpendicular to the $   u$--plane.  \end{thm}

\noindent The hodograph transformation   (\ref{eq:guidingcenter})  uncovers, in particular,  a hidden Coriolis--like force $ - ( \varepsilon ^3 /4) \, J \dot u $; this seems to be the first example when such a force arises in an \textit{inertial} reference frame.

\begin{figure}[thb]
	\center{  \includegraphics{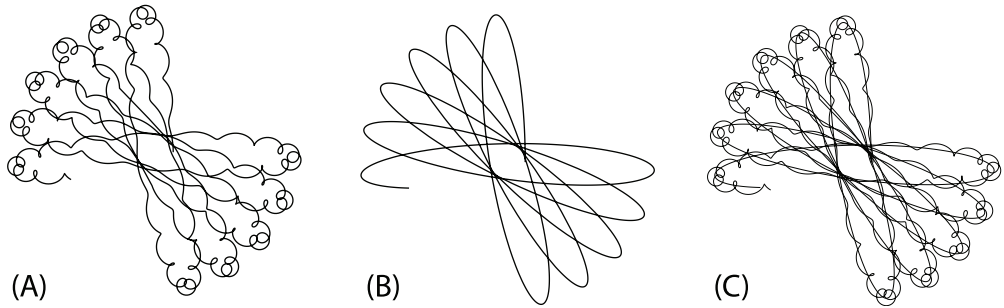}}
	\caption{A typical trajectory $ x  $; its guiding center $  u$; their superposition.}
	\label{fig:illustration1}
\end{figure}

  Figure~\ref{fig:illustration1}(C) shows some trajectories of   (\ref{eq:linear.trap})  with the motion of the guiding center superimposed on them. Some motions of the guiding center itself are illustrated in Figure~\ref{fig:illustration1}(B). It should be noted that the ``magnetic" effect
  $ \varepsilon ^3 J \dot  u /4$ is of higher order than the quadratic restoring force $ -\varepsilon ^2  u /4$; this explains why for
  $\varepsilon$ small the ``petals" become more closely spaced,  Figure~\ref{fig:illustration1}.
\vskip 0.1 in
Before proceeding with proof, we make a few observations that become clear in the course of the proof.
\vskip 0.1 in
\begin{rem} \rm Contrary to what one might expect, any averaging transformation {\it  must } involve $ \dot x   $. In other words, transforming just the configuration $ x  $ alone can not get rid of the time--dependence in the original equation {\it  in
principle}.  Put differently, {\it  the class of contact transformations is insufficiently wide to carry out the averaging of our system (\ref{eq:linear.trap})}. This is explained in the  Section \ref{impossibility}.
\end{rem}

\begin{rem}  \rm
The formal averaging procedure due to Kapitsa \cite{kapitsa} described in Landau--Lifshitz \cite{landau-lifshits} is incomplete and may give an incorrect result if one is not cautious or not lucky, as pointed out in  Section \ref{impossibility}.
\end{rem}

\begin{rem}  \rm  It is tempting to study the system in the rotating frame (as had been done, \cite{HB2005,THB2002}), since in such a frame the equations become autonomous, exchanging  the time dependence for the two inertial forces: the Coriolis force and the centrifugal force (in fact, the system in such coordinates is linear and thus admits explicit solutions, \cite{B1918,B1975,B1976,K2013,S01,CMMP2010,Ki11}). The motions we are studying here correspond to a fixed region of $ 0\in {\mathbb R}  ^4 $; in the rotating reference system this region is not fixed as $ \omega $ becomes large, which makes this frame poorly suited for using normal form. As a side remark, a physical manifestation of the dependence of the neighborhood on
$\omega$ is the fact that   solutions such as ones in Figure~\ref{fig:illustration1} undergo rapid rotations around the origin when viewed in a rotating frame.
\end{rem}

\section{Rotating saddle in physical applications.}
Before giving a proof of our statement,  we mention that time-periodic non-autonomous equations   (\ref{eq:linear.trap}), as well as their autonomous version in the rotating frame, arise in numerous applications across many seemingly unrelated branches of classical and modern physics \cite{K2013a,K2013,Ki11}. In celestial mechanics the rotating saddle equations describe linear stability of the triangular Lagrange libration points $L_4$ and $L_5$ in the restricted circular three-body problem \cite{BB94,A1970}. By this reason, the classical work by Gascheau of 1843 may be considered as the first one that established stability conditions for a particle on a rotating saddle \cite{G1843,AKN2006,S2008}. However, it was not until Brouwer, one of the authors of the fixed point theorem in topology, considered in 1918 stability of a heavy  particle on a rotating slippery surface  \cite{B1918,B1975,B1976} that the rotating saddle trap per se became an object for investigation.

Indeed, according to Earnshaw's theorem an electrostatic potential cannot have stable equilibria, i.e. minima, since such potentials are harmonic functions. The theorem does not apply, however, if the potential depends on time; in fact, the 1989 Nobel Prize in physics was awarded to W. Paul \cite{Paul1990} for his invention of the trap for suspending charged particles in an oscillating electric field.
Paul's idea was to stabilize the saddle by ``vibrating'' the electrostatic field, by analogy with the so--called Stephenson-Kapitsa pendulum \cite{kapitsa,landau-lifshits,stephenson,Yudovich1998,Levi1998,Levi1999} in which the upside--down equilibrium is stabilized by vibration of the pivot.
Brouwer \cite{B1918,B1975} explicitly demonstrated that, instead of vibration, the saddle can also be stabilized by rotation of the potential (in two dimensions). This effect is used, e.g., in quantum optics, in the theory of rotating radio-frequency ion traps \cite{HB2005}.

In plasma physics equations (\ref{eq:linear.trap}) appear in the modeling of a stellatron -- a high-current betatron with stellarator fields used for accelerating  electron beams in helical quadrupole magnetic fields \cite{P1970,S01,RMC1983,C1986}.  In atomic physics the stable triangular Lagrange points were produced in the Schr\"odinger-Lorentz atomic electron problem by applying a circularly polarized microwave field rotating in synchrony with an electron wave packet in a Rydberg atom \cite{BB94}. This has led to a first observation of a non--dispersing Bohr wave packet localized near the Lagrange point while circling the atomic nucleus indefinitely \cite{MGG2009}. Recently, the rotating saddle equations  (\ref{eq:linear.trap})   reappeared  in the study of confinement of massless Dirac particles, e.g. electrons in graphene \cite{N2013}. Even stability of a rotating flow of a  perfectly conducting ideal fluid in an azimuthal magnetic field possesses a mechanical analogy with the stability of Lagrange triangular equilibria and, consequently, with the gyroscopic stabilization on a rotating saddle \cite{OP1996}. Finally, we note that in mechanical engineering equations  (\ref{eq:linear.trap}) describe  stability of a mass mounted on a non-circular weightless rotating shaft subject to a constant axial compression force \cite{V95,I1988}.

\section{Derivation and proofs.}

\subsection{The guiding center transformation.}
In an attempt to bring   (\ref{eq:linear.trap})  to  a normal form, let us choose a new variable $x  _1\in {\mathbb R}  ^2 $ via
\begin{equation}
 x  =x  _1 + \frac{\varepsilon ^2 }{4} S(t/ \varepsilon) x  _1.
	\label{eq:cv1}
\end{equation}
\begin{small}
\begin{rem}  \label{rem:motivation}{\bf Heuristically,}  this transformation is suggested by the following reasoning: at a fixed position $x  $, the force $ S(t/ \varepsilon ) x   $
rotates in a circle with angular velocity $ 2 \omega $. If an otherwise free particle were subject to such a force, it would move in a circle if we subtract the drift,  with $  | Sx   | $ being the centripetal force, and thus given by
  $ | S x  |  = (2\omega) ^2 r $, where $r$ is the radius of the circle. With
$ \omega = 1/ \varepsilon $  this gives  $ r = \varepsilon ^2 | Sx  |/4$, explaining the choice (\ref{eq:cv1}).
\end{rem}
\end{small}
 \begin{thm} \label{guidingctr}
The transformation  (\ref{eq:cv1}) converts the system  (\ref{eq:linear.trap}) into the form
 \begin{equation}
	\ddot x  _1 -\varepsilon   SJ \dot x_1   +\frac{\varepsilon ^2}{4}x_1+
	\frac{\varepsilon ^3}{4} J \dot x_1 -
	 \frac{\varepsilon ^4}{16}S x_1 = O ( \varepsilon ^5).
	\label{eq:first.normal.form}
\end{equation}
\end{thm}

\begin{rem}  \rm The coefficient matrices $ SJ$ and $S$  in
       (\ref{eq:first.normal.form}) have zero average; this may suggest that the corresponding terms can be killed by some transformation  and yield the averaged equation
\[
	\ddot x_2 + \frac{\varepsilon ^2 }{4}   x_2  + \frac{\varepsilon ^3 }{4} J \dot x_2   =
	 O( \varepsilon ^5 ), 	
\] 	
a wrong result which looks almost exactly like  the correct equation (\ref{eq:mainaveq1}), except for the    sign in front of the Coriolis term. This  shows that such formal averaging of the coefficients is illegal.
\end{rem}
\vskip 0.1 in
\noindent{\bf Proof of Theorem \ref{guidingctr}} is a routine calculation which we give for the sake of completeness.
Substituting the derivatives of the transformation   (\ref{eq:cv1})
 \begin{equation}
    \begin{array}{l}
    \dot x  =  \dot x_1  + \frac{\varepsilon ^2}{4}   \dot S x_1+
    \frac{\varepsilon ^2}{4}  S\dot x_1   \\[3pt]
    \ddot x  = \ddot x_1 +  \frac{\varepsilon ^2}{4}( \ddot S x_1 + 2 \dot S \dot x_1  +
    S \ddot x_1 )
    \end{array}
	\label{eq:ddox}
\end{equation}
 into   (\ref{eq:linear.trap}) yields
\begin{equation}
	\biggl( I+ \frac{\varepsilon ^2}{4}S\biggl)   \ddot x_1 +
	\frac{1}{2} \varepsilon ^2 \dot S  \dot x_1   + \frac{\varepsilon ^2 }{4}    \ddot S x_1  	+
	S\biggl(x_1 + \frac{\varepsilon ^2 }{4} S  x_1\biggl)   =0.
	\label{eq:long}
\end{equation}
Note that    $S$    satisfies
\begin{equation}
	\boxed{S ^\prime(\tau) = -2S(\tau) J, \ \  S ^{\prime\prime} ( \tau ) = - 4S(\tau). }
	\label{eq:RSrelation}
\end{equation}
Using the notations $ \tau = t/ \varepsilon $,    $ {}^\prime =   \frac{d}{d\tau} $, we have
$ \frac{d}{dt} = \varepsilon ^{-1} \frac{d}{d\tau} $, so that
\[
	 \dot S = \varepsilon ^{-1} S ^\prime = - 2 \varepsilon ^{-1} SJ
\]
and
\[
	 \ddot S= \varepsilon ^{-2} S ^{\prime\prime} = - 4 \varepsilon ^{-2} S
\]
 according to   (\ref{eq:RSrelation}). Substitution into   (\ref{eq:long})  gives
 \[
	\biggl( I+ \frac{\varepsilon ^2}{4}S\biggl)   \ddot x_1 -   \varepsilon     S  J \dot x_1   -
	 \cancel{  S x_1} 	+ S\biggl( \cancel{ x_1} + \frac{\varepsilon ^2 }{4} S  x_1\biggl)   =0;
\]
canceling two terms as indicated above  and using $ S ^2 =I $ we get
\[
	\left(I+ \frac{\varepsilon ^2}{4}S \right) \ddot x_1  - \varepsilon   SJ \dot x_1  +
	  \frac{\varepsilon ^2 }{4}   x_1 =0.
\]
Multiplying by $ (I+ \frac{\varepsilon ^2}{4}S )  ^{-1} = I- \frac{\varepsilon ^2}{4}S  +O( \varepsilon ^4) $, we obtain
\begin{equation}
	\ddot x_1 - \varepsilon  \biggl( I- \frac{\varepsilon ^2}{4}S \biggl)   SJ \dot x_1  +  \frac{\varepsilon ^2 }{4}
	\biggl( I- \frac{\varepsilon ^2}{4}S \biggl)     x_1 =O (\varepsilon ^5).
	\label{eq:ddoy1}
\end{equation}
Using $ S ^2 = I $ again turns  (\ref{eq:ddoy1}) into   (\ref{eq:first.normal.form}), completing the proof of Theorem \ref{guidingctr}.
 \hfill $\diamondsuit$
 \vskip 0.1 in

 Before proceeding with the rest of the normal form reduction (section \ref{higher order reduction}), we pause to   make a few heuristic remarks in the next two sections.

 \section{A dead end transformation: Kapitsa's averaging.}\label{impossibility}
   It is now tempting to make further change of variables of the form
 \begin{equation}
	 x_1 = (I+ \varepsilon ^kH(t/ \varepsilon ) ) x_2, \  \  k \geq 1,
	\label{eq:configspacetransforms}
\end{equation}
 to eliminate the $ O( \varepsilon ) $--terms in   (\ref{eq:first.normal.form}). Perhaps surprisingly,     no such transformation  will work, as can   be checked by explicit substitution. This leads to a somewhat unfortunate but unavoidable conclusion:

\begin{thm} No transformation of the form  (\ref{eq:configspacetransforms})  can eliminate
$ O( \varepsilon ) $--terms in   (\ref{eq:first.normal.form}). In other words, the class of contact transformations is not rich enough to do the averaging in our system.
\end{thm}

We conclude:

{\it   In order to average out the time--dependent terms in the rotating saddle trap equation,  it is necessary to widen the class of transformations    (\ref{eq:configspacetransforms}) to include non--contact ones.  }
\vskip 0.1 in


\begin{rem}   Kapitsa \cite{kapitsa} described formal averaging of
the scalar equation
\[
	\ddot x + a(t/ \varepsilon ) f(x) = 0,  	
\]
see also \cite{landau-lifshits}. This heuristic procedure   applies to   (\ref{eq:linear.trap}) as well, although our equation   (\ref{eq:linear.trap})  has a matrix instead of the   scalar coefficient $a$. It must be pointed out that Kapitsa's' heuristic procedure gives only $ O( \varepsilon ^2) $--terms, but not the cubic ones; and it is these terms that are responsible for the ``precession" in Figure~\ref{fig:illustration1}.
 \end{rem}

  \section{Effective potential: a heuristic derivation.}
For a fixed location $x_0$,  the   force vector
\begin{equation}
	F _0(t) = F (x_0,t) = - S(t/ \varepsilon )  x_0
	\label{eq:forcecircle}
\end{equation}
rotates counterclockwise. {\it  If the force were independent of $ x$ near $x_0 $,}   then a  non--drifting particle would move in a circle, with $F _0(t) $, the centripetal force, related to the radius vector
$ r  = x-x_0 $  via
$ F =- (2\omega )^2 r  = - (\varepsilon ^2 /4)(x-x_0)$,  so  that
\begin{equation}
	x=x_0+ \frac{\varepsilon ^2 }{4} F =x_0+  \frac{\varepsilon ^2 }{4}Sx_0.
	\label{eq:circle}
\end{equation}
 Now so far we pretended that $ F  $ given by (\ref{eq:forcecircle}) does not depend on $x$.
  But   (\ref{eq:circle})  gives a better approximation than $ x_0 $, suggesting a better approximation for the force:
  \begin{equation}
	F _1(t) = -S \,\left(x_0+  \frac{\varepsilon ^2 }{4}Sx_0\right) = -Sx_0-\frac{\varepsilon ^2 }{4} x_0,
	\label{eq:shiftedcircle}
\end{equation}
 showing that in this  approximation (improving on  (\ref{eq:forcecircle})) $ F _1(t)$  travels in a {\it   shifted} circle, so that   the average
 \[
	\overline{F _1(t)}=-\frac{\varepsilon ^2 }{4} x_0;
\]
this is a restoring force (corresponding to the last term on the left--hand side in the averaged equations   (\ref{eq:mainaveq1})).

  To repeat the above in a more intuitive way, consider the force $ F =  S(t/ \varepsilon ) (x_0+r (t)) $ changing with $t$, as
    $ r (t)  $ travels in a small circle counterclockwise with the angular velocity $  2/ \varepsilon $. We have
    \[
	S(t/ \varepsilon ) (x_0+r (t))= Sx_0+Sr(t).
\]
The key point is that the last term is constant:
\begin{equation} F
	S(t/ \varepsilon) r (t) =r (0) ,
	\label{eq:Rr}
\end{equation}   thus providing the bias mentioned earlier. This constancy is due to the fact that the $t$--dependence in
$ S(t/ \varepsilon ) r (t) $ enters in both $S$ and $ r $ and the two dependencies happen to cancel.
  Indeed, let us temporarily fix $ S $, while allowing $r $ to change; then
$ S r  $ travels {\it  clockwise} since $S$ is orientation--reversing. But if we fix $r $ instead and let $S$  depend  on $t$, then $ S r $ would rotate  {\it  counterclockwise}  with the same angular velocity. The two rotations thus cancel,  explaining the constancy of
$ S r $.

 \section{Higher order reduction}
 \label{higher order reduction}
 As  noted in Section \ref{impossibility}, we must widen the class of transformations if we are to eliminate the leading time--dependent terms in    (\ref{eq:first.normal.form}).
 To that end we rewrite (\ref{eq:first.normal.form}) as a system
\[
   \left\{ \begin{array}{l}
    \dot x_1   =y _1 \\[3pt]
    \dot y _1  =  \varepsilon SJy _1- \frac{\varepsilon ^2 }{4}x_1- \frac{\varepsilon ^3 }{4}y _1+
    \frac{\varepsilon ^4 }{16} Sx  _1+ O(\varepsilon ^5),   \end{array} \right.
\]

or, more compactly as a system in $ {\mathbb R}  ^4 $:
\begin{equation}
	\dot z_1 =(A_0+ \varepsilon A_1+ \varepsilon ^2 A_2+ \varepsilon ^3 A_3+\varepsilon ^4 A_4+ O(\varepsilon ^5))\,  z_1,
	\label{eq:vectorform}
\end{equation}
 where
 \begin{equation}
	z_1= \left( \begin{array}{c} x  _1  \\ y _1 \end{array} \right), \  \
	A_0= \left( \begin{array}{cc} 0 & I \\ 0 & 0 \end{array} \right) , \  \
	A_1= \left( \begin{array}{cc} 0 & 0 \\ 0 & SJ \end{array} \right) , \  \
	\label{eq:Ak}
\end{equation}
  \begin{equation}
	A_2= - \frac{1}{4} \left( \begin{array}{cc} 0 & 0 \\ I & 0 \end{array} \right) , \  \
	A_3= - \frac{1}{4}\left( \begin{array}{cc} 0 & 0 \\ 0 &   J \end{array} \right), \ \
	A_4= \frac{1}{16} \left( \begin{array}{cc} 0 & 0 \\ S & 0 \end{array} \right).
	\label{eq:Ak}
\end{equation}

 \begin{thm}\label{thm:hodograph}
  The ``hodograph image"
\begin{equation}
	u = x_1- \frac{\varepsilon ^3 }{4}SJ \dot x  _1
	\label{eq:hodograph}
\end{equation}
of solutions of   (\ref{eq:first.normal.form})  satisfies
\begin{equation}
	\ddot  u - \frac{\varepsilon ^3}{4}    J\dot u+ \frac{\varepsilon ^2}{4} u = O( \varepsilon ^4),
	\label{eq:mainaveq}
\end{equation}
 i.e. behaves as a particle in the  potential $ V(u)  = \frac{\varepsilon ^2}{8} u^2 $   and subject to a magnetic force in the constant magnetic field $ B=\varepsilon ^3/4$ perpendicular to the $u$--plane.
\end{thm}
 Before giving the proof, we note that Theorem \ref{thm:hodograph} implies Theorem \ref{mainthm}. Indeed, from   (\ref{eq:cv1})
 we have
 \[
	x_1= x- \frac{\varepsilon ^2}{4} Sx+ O( \varepsilon ^4);
\]
substituting this into   (\ref{eq:hodograph})  gives
\begin{equation}
	 u = x  - \frac{\varepsilon ^2 }{4} S(x-  \varepsilon   J \dot x) + O( \varepsilon ^4) ,
	\label{eq:hodograph1}
\end{equation}
with the conclusion that this transformation converts the original equation (\ref{eq:linear.trap}) into     (\ref{eq:mainaveq}). This holds even if the
$O( \varepsilon ^4)$--term in   (\ref{eq:hodograph1})   is deleted, since it affects only the $O( \varepsilon ^4)$--terms in the equation for $u$. This proves Theorem \ref{mainthm} modulo Theorem \ref{thm:hodograph}.

\begin{rem}  The magnetic term in   (\ref{eq:mainaveq})  is of higher order than the restoring term. This is confirmed by the numerical evidence given in
Figure~\ref{fig:illustration1}: for small $\varepsilon$ the precession is slow compared to the oscillations.
\end{rem}

\noindent {\bf Proof of Theorem \ref{thm:hodograph}}.  This theorem can be verified by a direct calculation, simply by substituting   (\ref{eq:hodograph1})  into   (\ref{eq:mainaveq})  and using the governing equation
$ \ddot x   = - S x  $. This, however, would give no hint on how   (\ref{eq:hodograph1})  was discovered, and we proceed with a normal form reduction which led to   (\ref{eq:hodograph}) (and thus to   (\ref{eq:hodograph1})).

We wish to eliminate time--dependence of the coefficients of  our system (\ref{eq:vectorform}) up to and including
 order $ \varepsilon ^4 $.  We do so by the standard normal form argument.

 \paragraph {Averaging  the $ O( \varepsilon ) $--term.}   We seek to kill time--dependence in the $ O( \varepsilon ) $--term in (\ref{eq:vectorform}) via the change of variables
 \begin{equation}
	z_1=(I+\varepsilon ^2 T_1)z_2, \  \  T_1=T_1(t/ \varepsilon ),
	\label{eq:T1}
\end{equation}
where $T_1(\tau)$ is a periodic $ 4 \times 4 $ matrix function of period $   \pi $.
 Substituting this into   (\ref{eq:vectorform}) and using
 \[
	(I+\varepsilon ^2 T_1)^{-1} = I - \varepsilon ^2 T_1+ \varepsilon ^4 T_1 ^2 + O( \varepsilon ^6),
\]
 we obtain
 \begin{equation}
	\dot z_2 = (B_0+ \varepsilon B_1+ \varepsilon ^2 B_2+ \varepsilon ^3 B_3+\varepsilon ^4 B_4+ O(\varepsilon ^5)) z_2,
	\label{eq:firstsub}
\end{equation}
 where
 \begin{equation}
	B_0=A_0, \  \  B_1 = A_1-T_1 ^\prime,
	\label{eq:Bs}
\end{equation}
and
	\begin{equation}
    \begin{array}{l}
         B_2=A_2+[A_0,T_1] \\[3pt]
         B_3=A_3+[A_1,T_1]+T_1T_1 ^\prime \\[3pt]
	 B_4=A_4+[A_2,T_1]-T_1[A_0,T_1]   \end{array}
	\label{eq:Bs1}
\end{equation}
with brackets denoting commutator of matrices.
 Note that according to our notation $ T ^\prime = \varepsilon ^{-1} \dot T$, so that $ T ^\prime = O(1) $.
By setting
\begin{equation}
	T_1= - \frac{1}{2} \left( \begin{array}{cc} 0 & 0 \\ 0 & S \end{array} \right)
	\label{eq:T1formula}
\end{equation}
 we     get
 \[
	B_1=A_1-T_1 ^\prime = 0,
\]
 as follows from   (\ref{eq:Ak})  and   (\ref{eq:RSrelation}).  Substituting   (\ref{eq:T1formula}) into   (\ref{eq:Bs1})    we compute
 \begin{equation}
	B_2 =- \frac{1}{4} \left( \begin{array}{cc} 0 & 2S \\ I & 0 \end{array} \right), \  \
	B_3=  \frac{1}{4} \left( \begin{array}{cc} 0 & 0 \\ 0 & J \end{array} \right) , \  \
	B_4= -\frac{1}{16} \left( \begin{array}{cc} 0 & 0 \\ S & 0 \end{array} \right);
	\label{eq:Bs2}
\end{equation}
 summarizing, our equation becomes
 \begin{equation}
	\dot z_2 = (B_0+ \varepsilon ^2 B_2+ \varepsilon ^3 B_3+\varepsilon ^4 B_4+ O(\varepsilon ^5)) z_2.
	\label{eq:firstsub1}
\end{equation}

  \paragraph {Averaging of the $ \varepsilon ^2 $--term.} We now eliminate $t$ from the $ B_2 $ term in   (\ref{eq:firstsub1}) by seeking the transformation
  \begin{equation}
	z_2=(I+ \varepsilon ^3 T_2)z_3, \ \  T_2= T_2(t/ \varepsilon ),
	\label{eq:T2transformation}
\end{equation}
	where $ T_2 (\tau)$ is a  matrix function periodic in $\tau$ of period $ 2 \pi $.
Substitution of   (\ref{eq:T2transformation})  into  (\ref{eq:firstsub}) gives the new system
\[
	\dot z_3=M_2 z_3	
\]
where
\begin{equation}
	M_2= (I+ \varepsilon ^3 T_2) ^{-1} M_1 (I+ \varepsilon ^3 T_2)
	- (I+ \varepsilon ^3 T_2)^{-1} \varepsilon ^2 T_2 ^\prime .
	\label{eq:M2}
\end{equation}
 Note that we used the fact that $ \dot T_2 = \varepsilon ^{-1} T_2 ^\prime $. Multiplying  out (\ref{eq:M2})
 and collecting the like powers of $\varepsilon$ we obtain
\begin{equation}
	M_2= A_0+ \varepsilon ^2 (B_2-T_2 ^\prime)+ \varepsilon ^3 \ (B_3+[A_0, T_2])  + \varepsilon ^4B_4+
	O( \varepsilon ^5);	
	\label{eq:M2expanded}
\end{equation}
note that the $ \varepsilon ^4 $--term was unaffected by the transformation. To kill the $t$--dependence in the
$ \varepsilon^2 $--term we choose $ T_2$ so as to turn
$ B_2-T ^\prime $ into the average of $ B_2 $:
\begin{equation}
	B_2- T_2 ^\prime 	= \overline{B_2} = - \frac{1}{4} \left( \begin{array}{cc} 0 & 0 \\I & 0 \end{array} \right),
	\label{eq:B2average}
\end{equation}
 This condition, along with the requirement of periodicity, dictates the choice
 \begin{equation}
	T_2=-\frac{1}{4} \left( \begin{array}{cc} 0& SJ \\ 0 & 0 \end{array} \right) .
	\label{eq:T2choice}
\end{equation}
 Substituting this into   (\ref{eq:M2expanded}) yields
\[
	M_2=A_0+ \varepsilon ^2 \overline{B_2}+ \varepsilon ^3 B_3+ \varepsilon ^4 B_4+ O( \varepsilon ^5),
\]
 where we used the fact that $ B_3+[A_0,T_2]= B_3 $, since $ T_2 $ commutes with $ A_0 $.

 \paragraph {Reduction of the $ \varepsilon ^4 $--term.} Note that the cubic term turned out to be time--independent, and thus we need to average the quartic term. To that end we subject the system
 \[
	\dot z_3 = M_2 z_3	
\]
to the transformation
 \[
	z_3 = (I + \varepsilon ^5T_4) z_5	
\]
  with the periodic matrix function $ T_4$ chosen so as to kill time dependence in $ B_4 $.\footnote{We use the subscript $4$ for consistency, noting that   $ T_3=0 $, i.e. that the identity transformation is needed for the cubic terms.}
  The matrix
  \[
	M_3 = (I + \varepsilon ^5T_3)^{-1} M_2 (I + \varepsilon ^5T_3) - (I + \varepsilon ^5T_3) ^{-1}
	\varepsilon ^3 T_3 ^\prime
\]
of the transformed system differs from $ M_2 $ only in the terms starting with $ \varepsilon^4$:
\[
	M_3 = M_2- \varepsilon ^4 T_4^\prime ,
\]
 and thus we must choose $ T_4 $ so as to kill the time--dependence in the coefficient of $ \varepsilon ^4 $:
 \[
	T_4^\prime = B_4 =
	- \frac{1}{16} \left( \begin{array}{cc} 0 & 0 \\ S & 0 \end{array} \right) ,
\]
or
\begin{equation}
	T_4=\frac{1}{32}  \left( \begin{array}{cc} 0 & 0 \\ SJ & 0 \end{array} \right) .
	\label{eq:T4}
\end{equation}
 \vskip 0.3 in
Denoting $ z_5=w $, we obtain the averaged system
 \begin{equation}
	\dot w= (A_0 + \varepsilon ^2  \overline{B_2}+ \varepsilon ^3 B_3+ O( \varepsilon^5))\, w,
	\label{eq:reducedeq}
\end{equation}
	or, explicitly,
\begin{equation}
	\frac{d}{dt} \left( \begin{array}{c} u \\  v \end{array} \right) =
	\left( \begin{array}{cc} 0 & I \\ - \frac{\varepsilon ^2 }{4}I  &  \frac{\varepsilon ^3 }{4} J \end{array} \right)
	 \left( \begin{array}{c}  u \\  v \end{array} \right) + O( \varepsilon^5)
	 \left( \begin{array}{c}  u \\  v\end{array} \right) .
	\label{eq:reducedeq1}
\end{equation}
It follows that $ u$ satisfies
\begin{equation}
	\ddot  u- \frac{\varepsilon ^3 }{4} J\dot  u+ \frac{\varepsilon ^2 }{4}{u}= O (\varepsilon ^4);
	\label{eq:secondorder}
\end{equation}
indeed, according to the first equation in   (\ref{eq:reducedeq1})
\[
	\dot  u=  v+ O( \varepsilon ^5);
\]
differentiating this by $t$ gives
\[
	\ddot  u= \dot  v+ O( \varepsilon ^4)
\]
-- note the drop in the power of $\varepsilon$ due to differentiation (recall that $ d/dt= \varepsilon ^{-1} d/d\tau $). Substituting $ \dot  v $ from the second equation in   (\ref{eq:reducedeq1}) results in  (\ref{eq:secondorder}).

It remains to find the explicit form for the  averaging transformation
\[
	z_1= (I+ \varepsilon^5 T_4)(I+ \varepsilon^3 T_2)(I+ \varepsilon^2 T_1)w.
\]
Expanding the product in the powers of $\varepsilon$, we write the transformation as
 \begin{equation}
	  I + \varepsilon^2 T_1+ \varepsilon ^3 T_2+O(\varepsilon ^5);
	\label{eq:composion}
\end{equation}
  substituting the expressions  for $ T_1 $ and $ T_2 $ (see (\ref{eq:T1formula}) and  (\ref{eq:T2choice})) and reading off the first component,  we obtain (\ref{eq:hodograph}),  as claimed in the statement of  Theorem \ref{thm:hodograph}.
 \hfill $\diamondsuit$

%
%


\section{Conclusion}

We showed that the rapid rotation of the symmetric saddle potential creates a weak Lorentz--like,  or a Coriolis--like force, in addition to an effective stabilizing potential -- all in the inertial frame. As a result, the particle in the rotating saddle exhibits, in addition to oscillations caused by effective  restoring force,  a slow prograde precession in the inertial frame caused by this pseudo--Coriolis effect. By finding a hodograph--like ``guiding center" transformation using the method of normal form, we found the effective equations of this precession that coincide with the equations of the Foucault's pendulum \cite{Khein1993}.
Interpretation of the unconventional Coriolis--like force arising in the inertial frame in the spirit of the geometric magnetism \cite{BR1993,BS2010,BS2011} is an open problem and would be welcome.

\section{Acknowledgments}

Mark Levi gratefully acknowledges support by the NSF grant   DMS-0605878. Oleg Kirillov is thankful for partial support through the EU FP7 ERC grant ERC-2013-ADG-340561-INSTABILITIES.

\end{document}